\documentclass{PoS}

\title{Interplay of gaugino (co)annihilation processes in the context of a precise relic density calculation}

\ShortTitle{Interplay of gaugino (co)annihilation processes}

\author{Julia Harz\\
        Sorbonne Universit\'es, Institut Lagrange de Paris (ILP), 98 bis Boulevard Arago, 75014 Paris, France\\
        Sorbonne Universit\'es, UPMC Univ Paris 06, UMR 7589, LPTHE, F-75005, Paris, France\\
        CNRS, UMR 7589, LPTHE, F-75005, Paris, France}

\author{Björn Herrmann\\
LAPTh, Universit\'e Savoie Mont Blanc, CNRS, 9 Chemin de Bellevue, F-74941 Annecy-le-Vieux, France}

\author{Michael Klasen, Karol Kova\v{r}\'ik, \speaker{Patrick Steppeler}\\
        Institut für Theoretische Physik, Westfälische Wilhelms-Universität Münster, Wilhelm-Klemm-Stra\ss e 9, D-48149 Münster, Germany\\
        E-mail: \email{p\_step04@uni-muenster.de}}
        
\abstract{The latest Planck data allow one to determine the dark matter relic density with previously unparalleled precision. In order to achieve a comparable precision on the theory side, we have calculated the full $\mathcal{O}(\alpha_s)$ corrections to the most relevant annihilation and coannihilation processes for relic density calculations within the Minimal Supersymmetric Standard Model (MSSM). The interplay of these processes is discussed. The impact of the radiative corrections on the resulting relic density is found to be larger than the experimental uncertainty of the Planck data.}

\FullConference{18th International Conference From the Planck Scale to the Electroweak Scale \\
		 25-29 May 2015\\
		 Ioannina, Greece }

\newcommand{\beq}{\begin{equation}}
\newcommand{\eeq}{\end{equation}}
\newcommand{\bea}{\begin{eqnarray}}
\newcommand{\eea}{\end{eqnarray}}
\newcommand{\DMNLO}{{\tt DM@NLO}}
\newcommand{\MO}{{\tt micrOMEGAs}}
\newcommand{\DS}{{\tt DarkSUSY}}
\newcommand{\SPheno}{{\tt SPheno}}
\newcommand{\DRbar}{{$\overline{\mathrm{DR}}$}}

\begin{document}

\section{Introduction}

The nature and composition of dark matter is certainly one of the big puzzles of modern physics. It is quite remarkable that albeit we do not know what dark matter actually is, we can determine its amount quite well. This determination is based on the latest Planck data \cite{Planck} and on the cosmological standard model including six free parameters. In terms of this model the dark matter relic density is found to be
\beq
	\Omega_{\mathrm{CDM}}h^2 = 0.1198 \pm 0.0015,
	\label{Planck}
\eeq
where $h$ denotes the present Hubble expansion rate in units of 100 km s$^{-1}$ Mpc$^{-1}$.

Amongst several attempts to explain dark matter, weakly interacting massive particles (WIMPs) have gained much attention in the last decades. This popularity is based mainly on the fact that a hypothetical particle with a mass in the GeV range and weak interactions leads roughly to the observed relic density. A canonical example for a WIMP is the lightest neutralino $\tilde{\chi}_1^0$, which is the lightest supersymmetric particle (LSP) in many scenarios of the MSSM. 

Starting from the theory side, the neutralino relic density can be calculated via
\beq
\Omega_{\chi} h^2 = m_{\chi} n_{\chi} / \rho_{\rm crit}.
\eeq
Therein $m_\chi$ denotes the neutralino mass, $\rho_{\rm crit}$ the critical density of the universe and $n_\chi$ the neutralino number density, which can be obtained by solving the Boltzmann equation
\begin{equation}
	\frac{\mathrm{d}n_\chi}{\mathrm{d}t} = -3 H n_\chi 
		- \left\langle\sigma_{\mathrm{ann}}v\right\rangle \Big[ n_\chi^2 
		- \left( n_\chi^{\mathrm{eq}} \right)^2 \Big].
	\label{Boltzmann}
\end{equation}
The neutralino equilibrium density is denoted by $n_\chi^{\mathrm{eq}}$ and the Hubble parameter $H$. Of central interest for our project is the thermally averaged annihilation cross section $\left\langle\sigma_{\mathrm{ann}}v\right\rangle$. This quantity includes all possible annihilation and coannihilation processes, where supersymmetric (SUSY) particles transform into Standard Model (SM)\footnote{In the context of the MSSM, the SM is supplemented with an extra Higgs doublet.} particles. Coannihilation processes are relevant (and often even dominant) when the mass difference between the LSP and the next-to-lightest supersymmetric particle (NLSP) becomes small. Typical NLSPs in the MSSM are stops, other neutralinos or charginos.

In practice, the Boltzmann equation is usually solved numerically by using a publicly available code such as \MO\ \cite{micrOMEGAs} or \DS\ \cite{DarkSUSY}. However, both codes provide the thermally averaged annihilation cross section entering the Boltzmann equation only at an effective tree level. The idea of our project \DMNLO\ is to improve on this by including full $\mathcal{O}(\alpha_s)$ corrections. The following processes have been implemented at $\mathcal{O}(\alpha_s)$ so far \cite{DMNLO_AFunnel, DMNLO_mSUGRA, DMNLO_NUHM, DMNLO_ChiChi, DMNLO_Stop1, DMNLO_Stop2, DMNLO_Stopstop}:
\bea
\label{GauginoProcesses}
\mathrm{Gaugino\ (co)annihilation:} & \tilde{\chi}_i^0\tilde{\chi}_j^0\rightarrow q\bar{q} &\quad\tilde{\chi}_k^\pm\tilde{\chi}_i^0 \rightarrow q\bar{q}'  \quad\tilde{\chi}_k^\pm \tilde{\chi}_l^\pm \rightarrow q\bar{q}\\
\label{CoanniProcesses}
\mathrm{Neutralino-stop\ coannihilation:}&  \tilde{\chi}_1^0\tilde{t}_1\rightarrow qV & \quad\tilde{\chi}_1^0\tilde{t}_1\rightarrow qH  \ \  \quad\tilde{\chi}_1^0\tilde{t}_1\rightarrow tg\\
\label{StopProcesses}
\mathrm{Stop\ annihilation:} &  \tilde{t}_1\tilde{t}^*_1\rightarrow VV & \quad \tilde{t}_1\tilde{t}^*_1\rightarrow VH\ \ \  \quad \tilde{t}_1\tilde{t}^*_1\rightarrow HH
\eea
Note that the indices are kept completely general in case of the gaugino\footnote{By gauginos we understand neutralinos and charginos.} (co)annihilation, i.e. $\{i,j\} = \{1,2,3,4\}$, $\{k,l\} = \{1,2\}$, and $q = \{u,d,c,s,t,b\}$. Hence, we also include coannihilations of different gauginos and light quark final states. Furthermore $V = \{\gamma, Z^0, W^\pm\}$ and $H = \{h^0, H^0,A^0,H^\pm\}$.

In the following we highlight selected findings of our publications. We focus on gaugino (co)annihilation and stop annihilation. Other aspects of our project are reviewed in the original publications and further recent proceedings \cite{DMNLO_Proceedings}.

\section{Some technicalities}

In this section we briefly summarise some of the main technical aspects of our project. Details can be found in \cite{DMNLO_ChiChi, DMNLO_Stop1, DMNLO_Stop2, DMNLO_Stopstop}.

The loop diagrams give rise to ultraviolet (UV) and infrared (IR) divergences. Both kinds of divergences are regularised via dimensional reduction \cite{Siegel}, where we distinguish between UV and IR poles. The UV divergences are removed via renormalisation. We use a hybrid on-shell / \DRbar\ scheme. More precisely, the bottom quark mass $m_b$ and the trilinear couplings $A_t$ and $A_b$ are defined as \DRbar\ quantities, whereas the quark and squark masses $m_t, m_{\tilde{t}_1}, m_{\tilde{b}_1}, m_{\tilde{b}_2}$ are treated on-shell. The renormalisation scale is set to 1 TeV, agreeing with the SPA convention \cite{SPA}. According to the Kinoshita-Lee-Nauenberg theorem \cite{KLN}, the IR divergences cancel when adding up the corresponding $2\to 3$ processes including the emission of an additional gluon to the virtual corrections of the $2\to 2$ processes. However, to allow for a seperate numerical integration of the phase spaces, additional work has to be done. There are mainly two ideas to deal with this problem, the dipole subtraction \cite{Dipols} and the phase space slicing method \cite{PhaseSpace}. We use the first one for gaugino (co)annihilation and the second one for neutralino-stop coannihilation and stop annihilation. Further details like the concrete implementation of the running of $m_b$ and $\alpha_s$ or higher order corrections to the bottom Yukawa coupling can be found in our original papers quoted above.

\section{Numerical results}

To analyse the impact of the $\mathcal{O}(\alpha_s)$ corrections on the (co)annihilation cross sections and the resulting relic density, we numerically investigate typical scenarios within the phenomenological MSSM (pMSSM). The eleven input parameters of our pMSSM setup are listed in table \ref{ScenarioList} for three reference scenarios. These parameters are handed over to \SPheno\ \cite{SPheno} which calculates the needed MSSM spectrum, i.e. all the particle masses, mixing angles etc.

\begin{table}
\begin{center}
\footnotesize
	\begin{tabular}{|c|ccccccccccc|}
		\hline
			 & $\tan\beta$ & $\mu$ & $ m_A$ & $M_1$ & $ M_2$ & $ M_3$ & $ M_{\tilde{q}_{1,2}}$ & $ M_{\tilde{q}_3}$ & $ M_{\tilde{u}_3}$ & $ M_{\tilde{\ell}}$& $ A_t$ \\ 
			\hline 
			I & 13.4 & 1286.3 & 1592.9 & 731.0 & 766.0 & 1906.3 & 3252.6 & 1634.3 & 1054.4 & 3589.6 & -2792.3\\			
			II & 27.0 & 2650.8 & 1441.5 & 1300.0 & 1798.4 & 1744.8 & 2189.7 & 2095.3 & 1388.0 & 1815.5 & -4917.5\\		
			III &  5.8 & 2925.8 & 948.8 & 335.0 & 1954.1 & 1945.6 & 3215.1 & 1578.0 & 609.2 & 3263.9 & 3033.7\\			
			\hline
	\end{tabular}
	\caption{pMSSM input parameters for three selected reference scenarios. All parameters except $\tan\beta$ are given in GeV. }
	\label{ScenarioList}
\end{center}
\end{table}

We list the most important (co)annihilation channels of these three scenarios in table \ref{ScenarioChannels}. Scenario I is dominated by gaugino (co)annihilation into heavy quarks\footnote{We also found scenarios where light quark final states contribute in a sizeable way, see \cite{DMNLO_ChiChi}.}. Note that coannihilations with the second neutralino or the chargino are favoured over self-annihilation of the lightest neutralino. This is due to Higgs resonances as $m_{\tilde{\chi}^0_1} + m_{\tilde{\chi}^0_2} = 738.2 + 802.4\ \mathrm{GeV} = 1540.6\ \mathrm{GeV} \approx m_{A^0} = 1592.9\ \mathrm{GeV}$ and $m_{\tilde{\chi}^0_1} + m_{\tilde{\chi}_1^+} = 738.2 + 802.3 \ \mathrm{GeV} = 1540.5\ \mathrm{GeV}\approx m_{H^+} = 1595.1 \ \mathrm{GeV}$ in this scenario. The second scenario is dominated by stop annihilation processes into electroweak final states. This is caused by a small mass splitting between the lightest neutralino and the stop. More precisely we have $m_{\tilde{t}_1} - m_{\tilde{\chi}^0_1} = 1361.7 - 1306.3\ \mathrm{GeV} = 55.4\ \mathrm{GeV}$. As before, many final final states and corresponding processes contribute in parallel. In the last scenario, all three classes given in equations (\ref{GauginoProcesses}) - (\ref{StopProcesses}) occur. The dominant contributions stem from neutralino-stop coannihilation. 

In the next subsections we show selected numerical results for scenarios I and II. The third scenario will be investigated in great detail in our next publication.

\begin{table}
\begin{center}
	\footnotesize
	\begin{tabular}{|rl|cccc|}
		\hline
		 &  & ~~~~ I ~~~~ & ~~~~ II ~~~~ & ~~~~ III ~~~~ &  \\
		\hline
		$\tilde{\chi}^0_1 \tilde{\chi}^0_1 \to$ & $t\bar{t}$ & 2\% &  & 16\% & \\
		                                        & $b\bar{b}$ & 9\% &  &  & \\
		\hline
		$\tilde{\chi}^0_1 \tilde{\chi}^0_2 \to$ & $t\bar{t}$ &  3\% & &  & \\
		                                        & $b\bar{b}$ & 23\% &   &  & \\		
		\hline
		$\tilde{\chi}^0_1 \tilde{\chi}^{+}_1 \to$ & $t\bar{b}$ & 43\% & & & \\
		\hline
		$\tilde{\chi}^0_1 \tilde{t}_1 \to$ & $th^0$ & & 1\% & 23\% & \\
		                                   & $tg$ &  & 6\% & 23\%  &\\
		                                   & $tZ^0$ &  &  & 5\%& \\
		                                   & $bW^+$ &  &  & 11\% &  \\ 
		\hline
		$\tilde{t}_1 \tilde{t}^{*}_1 \to$ & $h^0h^0$ & & 12\% & 5\%  &\\
		                                        		& $h^0H^0$ &  & 11\% &  & \\
		                                        		& $Z^0A^0$ &  & 7\% &  & \\
		                                        		& $W^\pm H^\pm$ & &14\% &  & \\
		                                        		& $Z^0Z^0$ &  & 8\% & 2\% & \\
		                                        		& $W^\pm W^\pm$ &  & 13\% & 3\% & \\
		\hline
		\multicolumn{2}{|c|}{Total} & 80 \% & 72\% & 88\% & \\
		\hline
	\end{tabular}
	\caption{Most relevant (co-)annihilation channels of the three reference scenarios. Channels which contribute less than 1\% to the thermally averaged cross section and/or are not implemented in our code are not shown.}
	\label{ScenarioChannels}
	\end{center}
\end{table}

\subsection{Gaugino (co)annihilation}

\begin{figure}
        \includegraphics[width=0.49\textwidth]{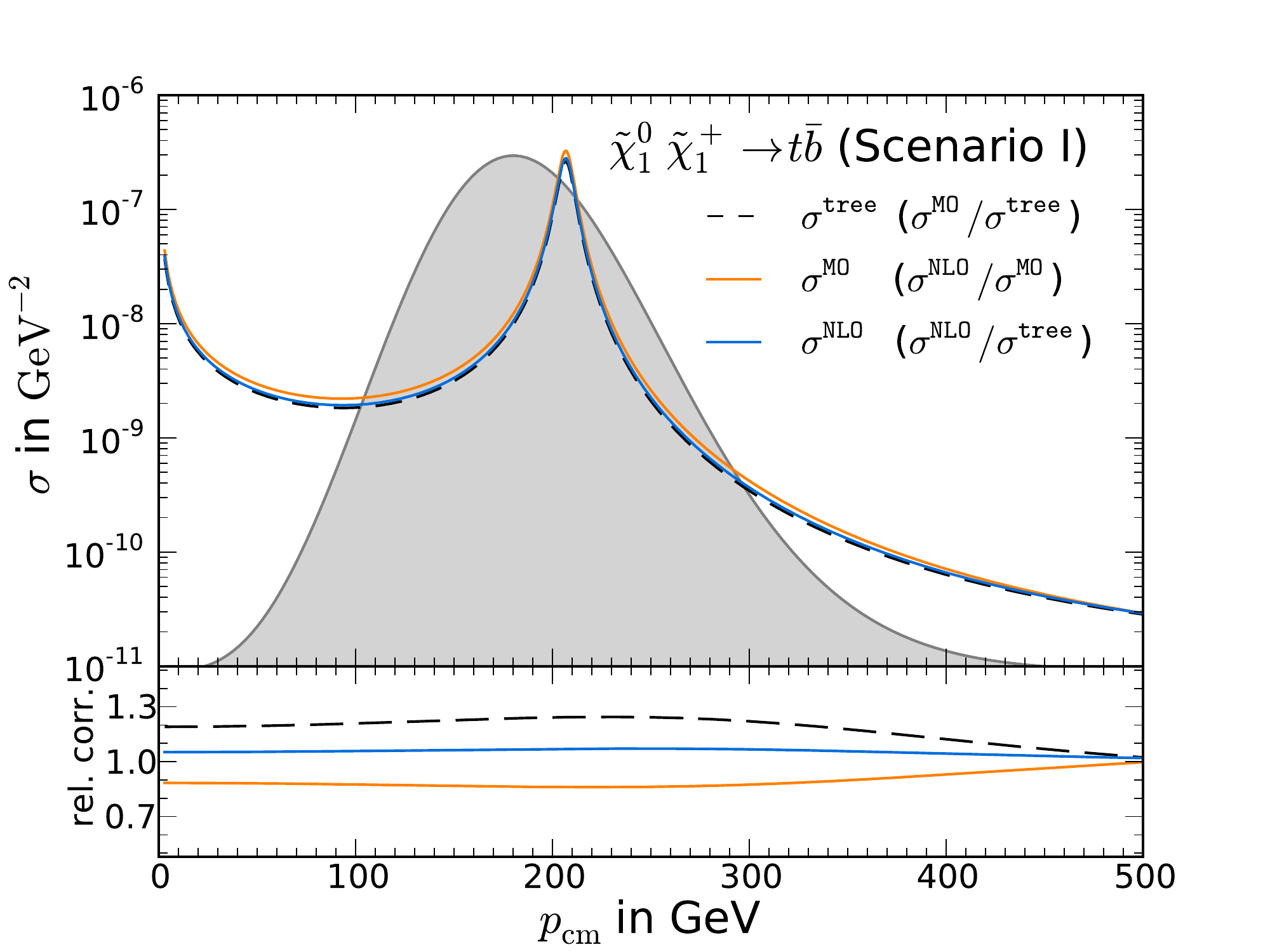}
        \includegraphics[width=0.49\textwidth]{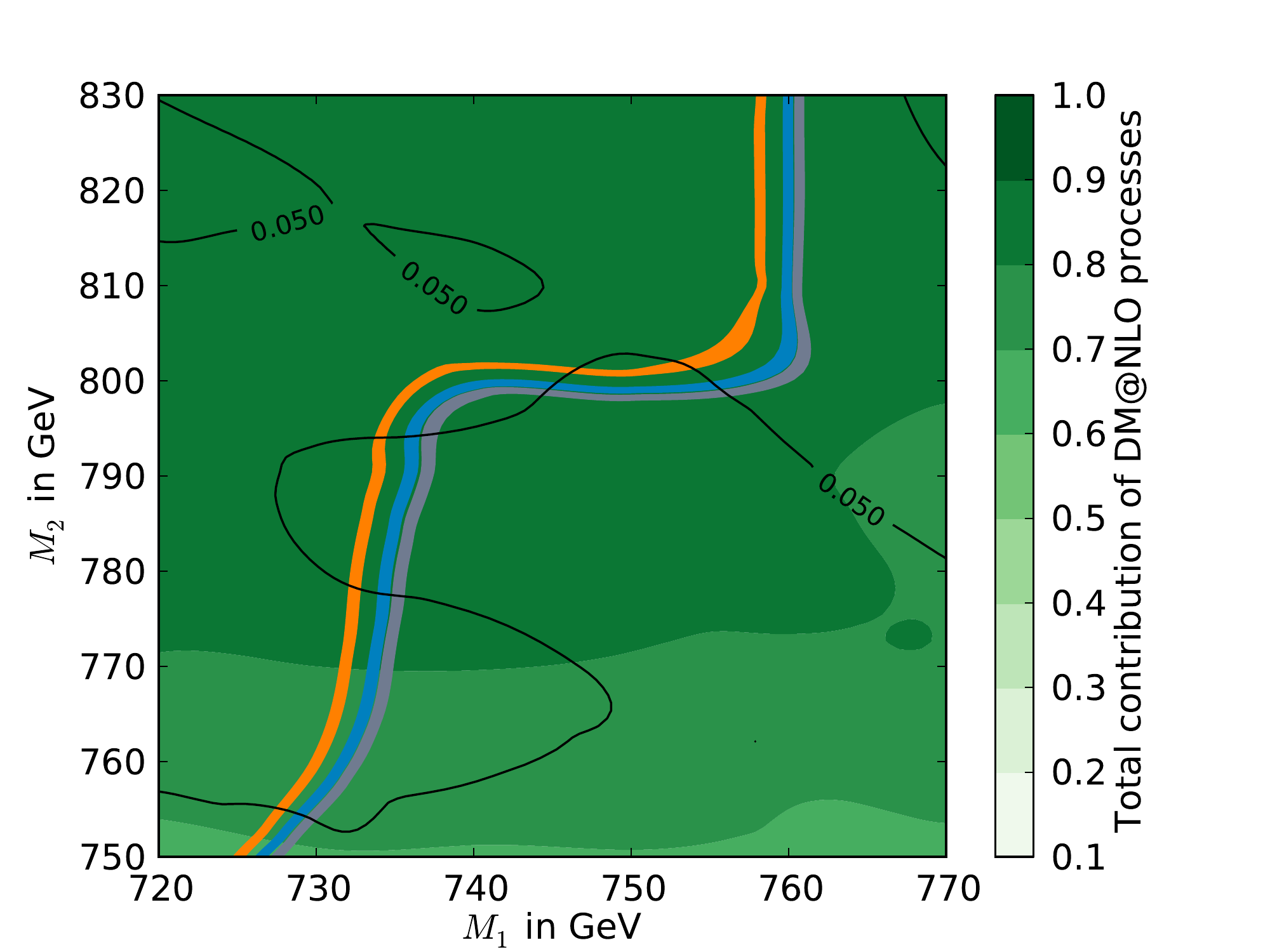}
\caption{Left: Tree level (black dashed line), \MO\ (orange solid line) and $\mathcal{O}(\alpha_s)$ corrections (blue solid line) for $\tilde{\chi}_1^0 \tilde{\chi}^+_1 \to t\bar{b}$. Right: Scan over the $M_1-M_2$-plane. The black contours depict the deviation between \MO\ and our full result.}
	\label{Fig:ChiChi2qq}
\end{figure}

On the left hand side of figure \ref{Fig:ChiChi2qq} the cross section of the process $\tilde{\chi}_1^0\tilde{\chi}_1^+\to t\bar{b}$ is depicted. This is the dominant channel of scenario I (cf. table \ref{ScenarioChannels}) and we recognise the aforementioned Higgs resonance at a center of mass momentum of $p_{\mathrm{cm}}\approx 200$ GeV. The grey background area illustrates the thermal distribution in arbitrary units. As this process peaks near the maximum of the thermal distribution, its relative contribution to the relic density calculation is large. The black dashed line denotes our tree level, the orange solid line the \MO\ effective tree level and the blue curve our full result including $\mathcal{O}(\alpha_s)$ corrections. The first two differ due to a different treatment of top and bottom masses and the use of effective couplings in the \MO\ code by up to 20~\%. However, our full result still differs from \MO\ by roughly 10-15~\% (see lower part of the plot). This propagates to the calculation of the relic density which is illustrated on the right hand side of figure \ref{Fig:ChiChi2qq}.

The green background colours denote the relative fraction of processes which our code \DMNLO\ supports at $\mathcal{O}(\alpha_s)$. This amounts to roughly 80~\% in the present case. The orange, grey and blue curves correspond to the part of the parameter space in the $M_1-M_2$ plane compatible with the Planck limits given in equation (\ref{Planck}) when using \MO, our tree level and our full next-to-leading order (NLO) result respectively. These three bands separate. The deviation between our full result and \MO\ amounts to $\sim$10~\%.

\subsection{Stop annihilation}

\begin{figure}
        \includegraphics[width=0.49\textwidth]{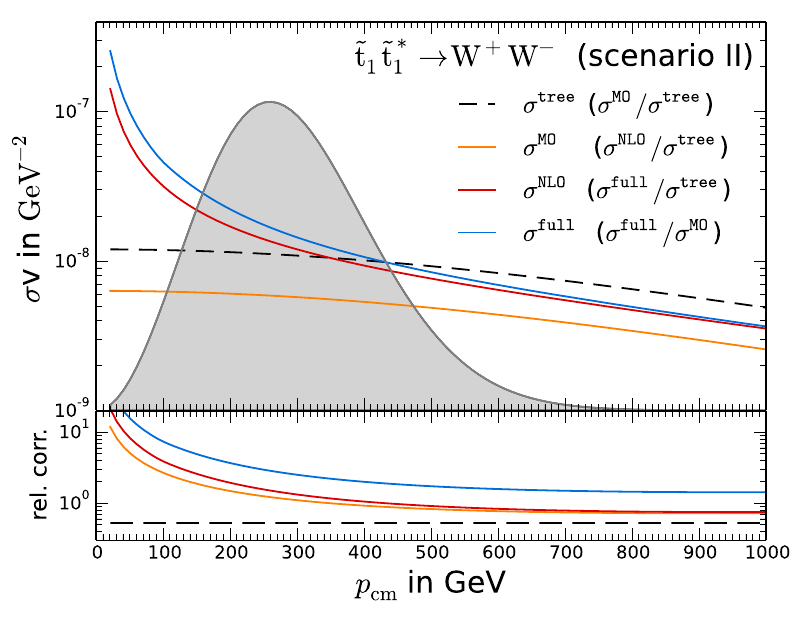}
        \includegraphics[width=0.49\textwidth]{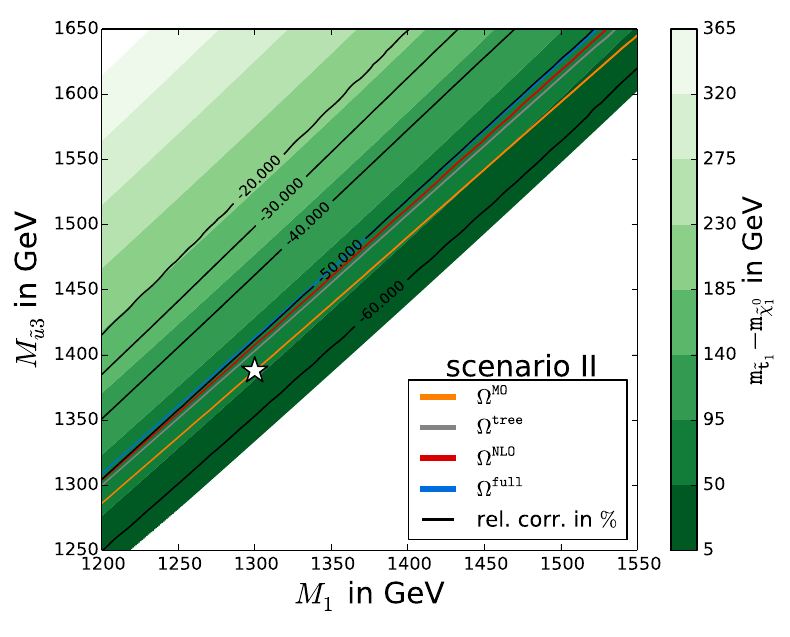}
\caption{Left: Tree level (black dashed line), \MO\ (orange solid line), $\mathcal{O}(\alpha_s)$ corrections (blue solid line) and full corrections (red solid line) for $\tilde{t}_1 \tilde{t}^*_1 \to W^+W^-$. Right: Scan over the $M_1-M_{\tilde{u}_3}$-plane. The black contours depict the deviation between \MO\ and our full result. The white star marks the precise position of scenario II.}
	\label{Fig:QQ2xx}
\end{figure}

We continue with the discussion of the process $\tilde{t}_1\tilde{t}_1^* \to W^+W^-$ which is amongst the most relevant ones of scenario II (cf. table \ref{ScenarioChannels}). The orange solid line on the left hand side of figure \ref{Fig:QQ2xx} represents the \MO\ cross section, which differs again from our tree level result shown as the dashed black line. This is due to a different treatment of the top quark mass. Note that these curves are monotonous, no resonances are present. The NLO result -- shown as the solid red curve -- diverges for small $p_{\mathrm{cm}}$ and corresponding small relative velocities $v$. This has a well-known origin: The incoming coloured particles can exchange $n$ gluons in a ladder-type diagram creating corrections proportional to $(\alpha_s/v)^n$, which is known as the Sommerfeld or Coulomb enhancement. For small $v$ perturbativity breaks down and these corrections have to be resummed to all orders to allow for a reliable prediction of the cross section. The precise implementation of this effect into our code is described in \cite{DMNLO_Stopstop}. The solid blue line denotes our full result and also incorporates the Coulomb contributions beyond $\mathcal{O}(\alpha_s)$. As expected, the red and blue line converge for large $p_{\mathrm{cm}}$. However, we would like to stress the impact of these corrections: For small $p_{\mathrm{cm}}$ our full calculation differs more than 100~\% from \MO.

As one might have guessed, these radiative corrections influence the resulting relic density quite heavily. This is illustrated on the right hand side of figure \ref{Fig:QQ2xx}. The orange, grey, red and blue lines correspond to the part of the $M_1-M_{\tilde{u}_3}$ plane compatible with the Planck limits. As it was the case for gaugino (co)annihilation, all lines separate. However, the separation between our tree level result and \MO\ is much bigger than previously and increases even further when including the NLO and Sommerfeld corrections. More precisely, our full result for the relic density differs more than 50~\% from \MO\ as indicated by the black contour lines.

\section{Conclusion}

The neutralino relic density is determined by annihilation and (co)annihilation cross sections of processes transforming supersymmetric particles into standard model ones. We have calculated the full $\mathcal{O}(\alpha_s)$ corrections to gaugino (co)annihilation, neutralino-stop coannihilation and stop annihilation as listed in equations (\ref{GauginoProcesses}) - (\ref{StopProcesses}). We found that in many pMSSM scenarios several of these processes occur in parallel. The impact of the radiative corrections on the resulting relic density is found to be larger than the experimental uncertainty given by the Planck data. Hence, these corrections should be taken into account when predicting the neutralino relic density or when extracting SUSY parameters from cosmological measurements.

\section*{Acknowledgements}
The speaker would like to thank the organisers of the PLANCK conference 2015 for the opportunity to present our work and to contribute to the proceedings. The pleasant atmosphere enabled several fruitful discussions. Special thanks go to Andreas Crivellin who took the speaker to the airport of Thessaloniki via car and saved him from being completely stuck in Ioannina. The work of P. S. is supported by the Deutsche Forschungsgemeinschaft, Project No. KL1266/5-1.

\end{document}